\documentclass[amssymb, nobibnotes,longbibliography,preprint]{revtex4-2}

\usepackage{graphicx,color,amsmath}
\usepackage[normalem]{ulem}

\newcommand{\insertplot}[5]{\begin{figure}
		\hfill\hbox to 0.05in{\vbox to #5in{\vfill
				\inputplot{#1}{#4}{#5}}\hfill}
		\hfill\vspace{-.1in}
		\caption{#2}\label{#3}
\end{figure}}
\newcommand{\inputplot}[3]{
	\special{ps: plotfile #1}
}
\newcounter{fig}

\newcommand{\ee}{\end{equation}}
\newcommand{\eea}{\end{eqnarray}}
\newcommand{\be}{\begin{equation}}
\newcommand{\bea}{\begin{eqnarray}}

\begin{document}
	
	\title{$Z_N$-balls: Solitons from $Z_N$-symmetric scalar field theory}
	\author{Fabien Buisseret$^{1,2}$}
	\email{fabien.buisseret@umons.ac.be}
	\author{Yves Brihaye$^3$}
	\email{yves.brihaye@umons.ac.be}
		\affiliation{
		$^1$ Service de Physique Nucl\'{e}aire et Subnucl\'{e}aire, Universit\'{e} de Mons, UMONS Research Institute for Complex Systems, 20 Place du Parc, 7000 Mons, Belgium. \\ $^2$ CeREF Technique, Chaussée de Binche 159, 7000 Mons, Belgium. \\
		$^3$ Service de Physique de l'Univers, Champs et Gravitation, Universit\'{e} de Mons, UMONS  Research Institute for Complex Systems,  20 Place du Parc, 7000 Mons, Belgium.
	}%

	\date{\today}
	\begin{abstract}
	We discuss the conditions under which static, finite-energy, configurations of a complex scalar field $\phi$ with constant phase and spherically symmetric norm exist in a potential of the form $V(\phi^*\phi, \phi^N+\phi^{*N})$ with $N\in\mathbb{N}$ and $N\geq2$, i.e. a potential with a $Z_N$-symmetry. Such configurations are called $Z_N$-balls. We build explicit solutions in $(3+1)$-dimensions from a model mimicking effective field theories based on the Polyakov loop in finite-temperature SU($N$) Yang-Mills theory. We find $Z_N$-balls for $N=$3, 4, 6, 8, 10 and show that only static solutions with zero radial node exist for $N$ odd, while solutions with radial nodes may exist for $N$ even. 
	\end{abstract}

\maketitle
\newpage

\section{\label{sec:intro}Introduction}

Q-balls, originally introduced in \cite{Coleman:1985ki}, are non-topological solitons with finite energy and size, typically obtained by solving the equations of motion of a complex Klein-Gordon field with U(1) invariance: ${\cal L}=\eta_{\mu\nu}\partial^\mu\phi\ \partial^\nu\phi^*-v(\phi^*\phi)$ with $\eta={\rm diag}(+---)$.

The U(1) symmetry allows for the ansatz
\be \label{qba}
\phi({\rm x})=f(r)\, {\rm e}^{i\omega\, t},
\ee
with ${\rm x}$ the position 4-vector, $r$ the radial coordinate and $\omega\in \mathbb{R}$. Boundary conditions $f(0)\neq0$ and $f(r \to \infty)=0$ are chosen. The equations of motion then reduce to 
\be\label{eomQ}
\Delta f+\omega^2 f-\frac{1}{2}\partial_f v(f^2)=0.
\ee
 This field equation allows for solitons called Q-balls in the absence of coupling to gravity, see \textit{e.g.} \cite{Coleman:1985ki,Lee:1991ax,Kusenko:1997ad,Volkov:2002aj,Nugaev:2016wyt,Nugaev:2019vru} provided that a non quadratic part is present in $v$. The potential is often chosen to be of the form $v(\phi^*\phi) = m^2 \phi^* \phi - \lambda_4 (\phi^* \phi)^2 + \lambda_6 (\phi^* \phi)^3$ with $\lambda_4$ and $\lambda_6>0$. Due to the U(1) symmetry of the Lagrangian, Q-balls have a charge $Q\sim\omega$. The solitons are called boson star in the presence 
of gravity, see \textit{e.g.} the review \cite{Schunck:2003kk}. More general solutions than (\ref{qba}) can be found in Minkowski spacetime for U(1)-symmetric potentials:
\begin{itemize}
	\item Spinning Q-balls: $\phi({\rm x})=f(r,\theta)\, {\rm e}^{i\omega\, t+iJ\varphi}$ \cite{Volkov:2002aj}, where $(r,\theta,\varphi)$ are the spherical coordinates. \\
\item Multipolar solutions:  $\phi({\rm x})=f(r,\theta,\varphi)\, {\rm e}^{i\omega\, t}$ \cite{Herdeiro:2020kvf}.\\
\item Coupled Q-balls \cite{Alcubierre:2018ahf,Brihaye:2009yr,Loiko:2018mhb}.
\end{itemize}

 In the present paper we propose to break the continuous U(1) symmetry of potential $v(\phi^*\phi)$ by allowing for terms depending on $\phi^N+\phi^{*N}$ with an integer number $N\geq 2$, i.e. by studying a complex scalar field theory with $
Z_N$-symmetry. To our knowledge, the influence of such a breaking of U(1) symmetry on the solutions has been poorly studied in general, but it is worth quoting the Montonen-Sarker-Trullinger-Bishop (MSTB) model \cite{Montonen:1976yk,SUBBASWAMY1981379,SARKER1976255,Rajaraman:1978kd,Alonso-Izquierdo:2018uuj} in which the potential can be written under the form $(1-\phi^*\phi)^2+\frac{\kappa}{2}\phi^*\phi+\frac{\kappa}{4}(\phi^2+\phi^{*2})$ with $\kappa\in\mathbb{R}$. The solutions of this $Z_2$-symmetric model have been well studied in $(1+1)$-dimensions, including generalizations of the potential, see e.g. the recent work \cite{Mandal:2018dok}. Here the potential under study will be of the form $V(\phi^*\phi, \phi^N+\phi^{*N})$, and the field equations will be formulated in $(3+1)$-dimensions. As argued in the next paragraph, such a form may be of interest in the field on finite-temperature gauge theories.

Solutions built from $Z_3$-symmetric models have indeed been studied in several works, see e.g.  \cite{Gupta:2010pp,Jin:2015goa,Biswal:2019xju,Kannika:2017ekw,Brihaye:2020rrp}, in the context of an effective model of SU(3) Yang-Mills theory at nonzero temperature $T$. In the latter case, the scalar field $\phi$ is identified with the mean value of the Polyakov loop, defined as $\phi(T, \vec x)=\left\langle  P\, {\rm e}^{i\, g\int^{1/T}_0d\tau A_0(\tau, \vec x)}\right\rangle $, with $A_0$ the temporal component of the Yang-Mills field and $(\tau, \vec x)$ the temporal and spatial coordinates respectively. $P$ is the path-ordering, $g$ is the strong coupling constant and units where $\hbar = c =k_B= 1$ are used. The gauge group SU($N$) may be considered to better outline our motivations. On the one hand, under gauge transformations belonging to the centre of the gauge group, i.e. $Z_N$, the Polyakov loop is multiplied by an overall factor. The $Z_N$ symmetry of Yang-Mills theory is therefore broken (present) if $\phi\neq 0$ ($=0$). On the other hand, it is known that $\phi\neq 0$ ($=0$) above (below) the deconfinement temperature $T_c$~\cite{Susskind:1979up,Weiss:1981ev,KorthalsAltes:1999xb}. Hence it has been conjectured that the confinement/deconfinement phase transition might be driven by the breaking of a global Z$_{N}$ symmetry~\cite{Yaffe:1982qf,Svetitsky:1982gs}. In that framework, the shape $V(\phi^*\phi, \phi^N+\phi^{*N})$ should be typical of any effective model of Yang-Mills theory based on the Polyakov loop \cite{Sannino:2005sk}. It is worth pointing out that such effective models are able to accurately reproduce the Yang-Mills equation of state provided potential parameters are temperature-dependent and fitted on lattice QCD data \cite{Ratti:2005jh,Ratti:2006wg}. Effective models are still worth of interest nowadays, e.g. in glueball dark matter models \cite{Halverson:2016nfq,Carenza:2022pjd}. 

Our work is structured as follows. In Sec. \ref{sec:def} we discuss the existence of static, spherically symmetric, solitons with finite energy in generic  $Z_N$-symmetric potentials. Such solitons are called ``$Z_N$-balls" in the following. Then in Sec. \ref{sec:pot} we propose a model mimicking features of Yang-Mills theory above deconfinement temperature. Numerical solutions are constructed for several values of $N$ in Sec. \ref{sec:num}. After a discussion of solutions going beyond $Z_N$-balls in Sec.\ref{sec:solup}, concluding comments, including a remark on solutions' stability, are given in Sec. \ref{sec:summary}.

\section{\label{sec:def}$Z_N$-balls: Definition and existence}

As proposed in the Introduction, let us consider a complex Klein-Gordon field $\phi$ with $Z_N$-symmetric potential $V(\phi^*\phi, \phi^N+\phi^{*N})$, such that $N\in\mathbb{N}$ and $N\geq2$. The Lagrangian reads 
\be\label{lag0}
{\cal L}=\partial_\mu\phi\ \partial^\mu\phi^*-V(\phi^*\phi, \phi^N+\phi^{*N}).
\ee
Using a polar decomposition of the complex field, i.e. $\phi = f {\rm e}^{i \delta}$, the equations of motion
can be set in the form
\begin{subequations}\label{eom1}
\begin{eqnarray}
\Box f-f\partial_\mu\delta\, \partial^\mu\delta+V^{(1,0)}\, f+V^{(0,1)}\, N\, \cos(N\delta)\, f^{N-1}&=0,\label{eom1a}\\
f\, \Box\delta +2\partial_\mu f\partial^\mu\delta-V^{(0,1)}\, N\, \sin(N\delta) f^{N-1}&=0, \label{eom1b}
\end{eqnarray}
\end{subequations}
with $V^{(1,0)}=\left.\partial_a V(a, 2f^N\cos(N\delta))\right|_{a=f^2}$ and $V^{(0,1)}=\left.\partial_a V(f^2, a)\right|_{a=2f^N\cos(N\delta)}$. 

Setting $\delta=\frac{2\pi k}{N}$ with $k\in\mathbb{Z}$, i.e.
\be\label{zna}
\phi=\phi_k({\rm x})=f({\rm x})\ {\rm e}^{i\frac{2\pi k}{N}}\quad {\rm with}\quad  k\in\mathbb{Z},\quad f\in\mathbb{R},
\ee
reduces the equations (\ref{eom1}) to a single equation for $f$, reading 
\be\label{eom2}
\Box f+V^{(1,0)}\, f+V^{(0,1)}\, N\, f^{N-1}=0.
\ee

We define the $Z_N$-ball ansatz as a solution with spherical symmetry, that is
\be\label{znp}
\phi=\phi_k(r)=f(r)\ {\rm e}^{i\frac{2\pi k}{N}},
\ee
with $r$ the radial coordinate. The conditions for a localized, regular, solution imply the following boundary conditions
\be\label{boundary}
f(0) = f_0>0 \ \ , \ \ f'(0) = 0, \ \  \ \ f(r \to \infty) \sim \frac{{\rm e}^{-mr}}{r} \to 0.
\ee
$m$ is the mass term in the Lagrangian: $V(\phi^*\phi, \phi^N+\phi^{*N})\sim m^2\phi^*\phi$ in a power expansion in $\phi$. The constraint $f_0 > 0$ is necessary to avoid the trivial solution $f(r)=0$ and Eq. (\ref{eom2}) with the ansatz (\ref{znp}) reads
\begin{equation}\label{effpot}
f''+\frac{2}{r}f'-\partial_f U(f)=0,
\quad {\rm with}\quad U(f)=\frac{1}{2}V(f^2,2f^N).
\end{equation}
By comparison with Eq. (\ref{eomQ}), one sees that the $Z_N$-ball equation (\ref{effpot}) is equivalent to that of a Q-ball with $\omega=0$ and potential $U(f)$. According to the study \cite{Volkov:2002aj}, $\omega=0$ Q-balls, hence $Z_N$-balls in our study, may exist if 
\be\label{existence}
U''(0)>0\quad {\rm and}\quad \min_f \left(\frac{2U(f)}{f^2}\right)<0.
\ee
In the $Z_N$-ball case, no continuum of solutions (parametrized by $\omega$) is possible, basically because the phase of $\phi_k$ is ``locked", see Eq. (\ref{zna}). This phase-locking also forbids spinning $Z_N$-balls, whose phase should be proportional to ${\rm e}^{iJ\varphi}$. For the same reason, multipole solutions as those studied in \cite{Herdeiro:2020kvf} are not expected. The best that can be found beyond the ground state is a discrete set of solutions distinguished by the phase $k=0,\dots,N-1$ in (\ref{zna}) and, if radially excited states exist, by the number of nodes of the scalar field.

Radially excited $Z_N$-balls may exist but not necessarily. The classical particle analogy sheds light on their existence. We recall indeed that Eq. (\ref{effpot}) can be interpreted as Newton's equation for a particle whose position if $f$ and for which $r$ plays the role of time in the potential $-U(f)$. This particle experiences a friction term $\frac{2}{r}f'$. The interested reader will find a detailed description of this classical analogy for Q-balls, Q-holes and Q-bulges in Ref. \cite{Nugaev:2019vru}. Figure~1 shows typical classical trajectories in effective potentials with different $N$. When $N$ is even, $U(f)=U(-f)$: Radially excited states a priori exist since the particle may oscillate between the two maxima before reaching $f=0$. When $N$ is odd, $U(f)$ may have no definite parity, and potentials can be found where the particle cannot reach an equilibrium position in $f=0$. In this case, radially excited states will not exist. In \cite{Brihaye:2020rrp} for example, only a solution with zero node was found for a $Z_3$-symmetric potential. It is worth mentioning that potentials with shapes similar to the presented odd-$N$ case are also found in some hairy black hole models \cite{Gubser:2005ih}.

\begin{figure}\label{fig:effpot}
	\centering
	\includegraphics[width=0.5\linewidth]{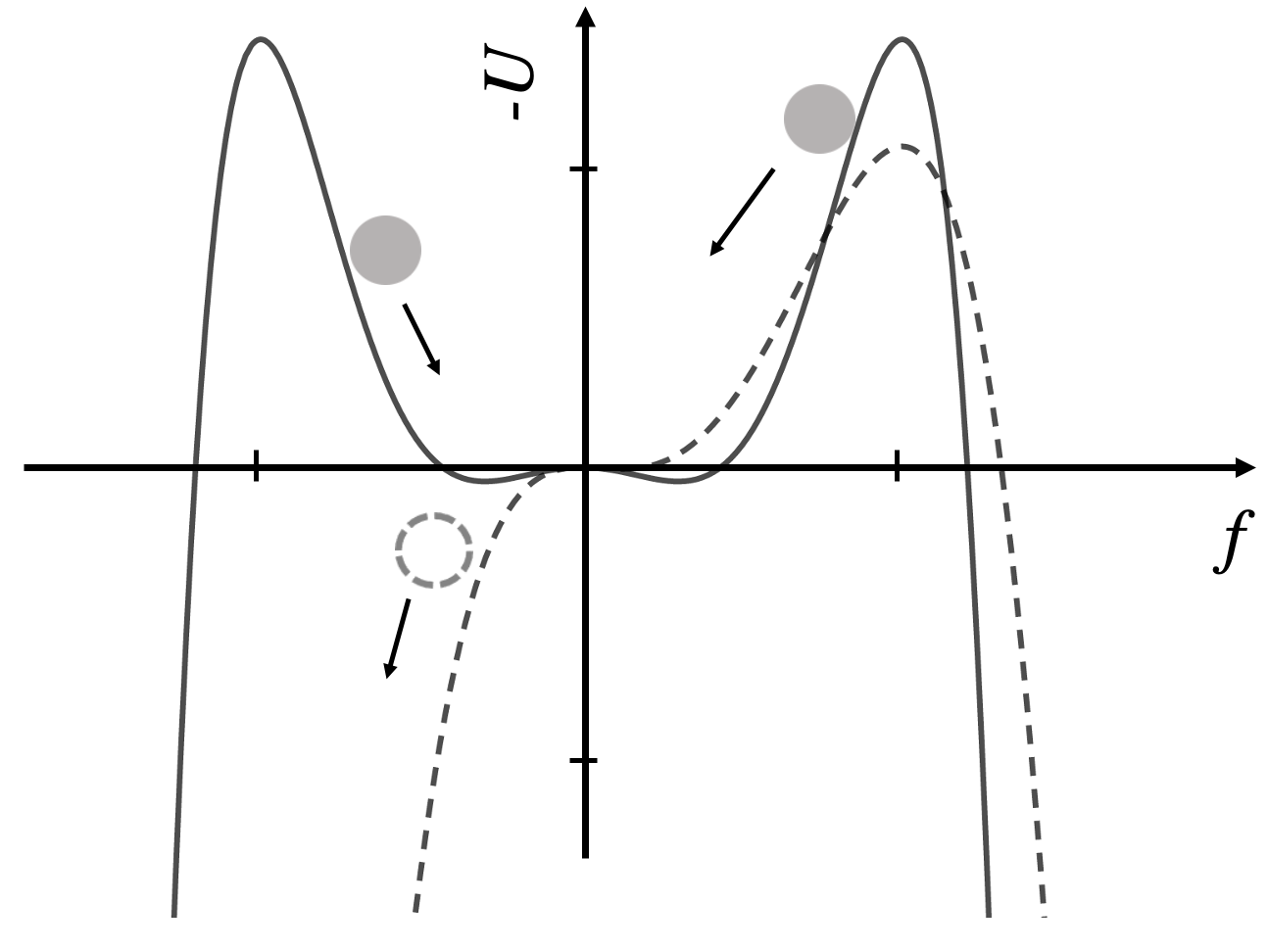}
	\caption{Mechanical analogy in $(3+1)$-dimensions for $Z_N$-balls with the ansatz (\ref{znp}) and the effective potential (\ref{effpot}). Typical plots corresponding to $N$ even (sold line) and odd (dashed line) are shown, and arrows indicate the trajectory of the equivalent classical particle. More precisely, the case $N$ even is obtained by plotting (\ref{potential_N4}) and the case $N$ odd is obtained by plotting (\ref{potential_N3}), with the horizontal axis in units of $l$ and the vertical axis in units of $m^2l^2/2$.}
\end{figure}

All the $Z_N$-symmetric potentials reducing to the same effective potential on the real axis will have the same radial profiles. As an example we consider the $N=4$ potential
\be
V=m^2 \phi^*\phi-\frac{\gamma}{2}(\phi^4+\phi^{*4})+ (\phi^*\phi)^3
\ee
and the $N=6$ potential
\be
V=m^2 \phi^*\phi-\gamma (\phi^*\phi)^2+\frac{1}{2}(\phi^6+\phi^{*6}).
\ee
Both are equivalent to a standard U(1)-symmetric potential $m^2 \phi^*\phi-\gamma (\phi^*\phi)^2+(\phi^*\phi)^3$ on the real axis. The latter potential may admit $\omega=0$ Q-balls: The effective potential reads $U(f)=m^2f^2-\gamma\, f^4+f^6$, and criterion~(\ref{existence}) states that $\omega=0$ Q-balls, hence $Z_4$-balls and $Z_6$-balls, exist for $\gamma^2>4m^2$. 

In the following, solitons will be further characterized by their mass $M$ and mean radius~$\left\langle R\right\rangle $~: 
\be
M = 4 \pi \int_0^{\infty} \epsilon dr ,\quad \epsilon(r)=f'^2 + U(f) \ ,
\ee
\be
\left\langle R\right\rangle = \frac{M_1}{M_0} \ \ , \ \ M_j = \int_0^{\infty} f^2(r) r^{2+j} dr .
\ee
In the above equations $\epsilon(r)$ is the local energy density. 

\section{\label{sec:pot}The model}

SU($N$) Yang-Mills theories at finite temperature are a field where $Z_N$-symmetric effective theories seem particularly relevant. We therefore propose a potential mimicking key features of such effective theories and numerically solve the field equation for the latter potential. We previously made a proposal of U(1)-symmetric potential focusing on the $N\to\infty $ limit of Yang-Mills theories \cite{Brihaye:2012uw}; here we propose a potential relevant for finite values, i.e. $N\geq 3$. We refer the reader to the aforementioned MSTB model in the case $N=2$.

The potential under study will be the $Z_N$-symmetric power-law potential
\begin{eqnarray}\label{potential1}
V(\phi^*\phi,\phi^N+\phi^{*N})&=\frac{m^2}{2}\left[\phi^*\phi-(1+\beta)(ml)^{2-N}\left( \phi^N+\phi^{*N}\right) +\frac{(ml)^{4-2N}}{N-1}\left( N(1+\beta)-1\right)(\phi^*\phi)^{N-1}  \right],\nonumber\\
& {\rm with}\quad m,\, \beta,\, l\in\mathbb{R}^+_0,\qquad N\in\mathbb{N},\qquad N\geq 3.
\end{eqnarray}
$m$ has the dimension of an energy and $\beta$, $l$ are dimensionless. According to Eq. (\ref{effpot}), it corresponds to the effective potential
\be \label{potential2}
U(f)=\frac{m^2}{2}\left[f^2-2(1+\beta)(ml)^{2-N}f^N +\frac{(ml)^{4-2N}}{N-1}\left( N(1+\beta)-1\right)f^{2N-2}  \right].
\ee
The latter potential is such that 
\begin{itemize}
	\item $f=0$ is a local minimum: $U(0)=U'(0)=0$ and $U''(0)=m^2>0$;\\
	\item $f=ml$ is a global minimum: $U(ml)=\frac{m^4l^2}{2}\beta\frac{2-N}{N-1}<0$, $U'(ml)=0$ and $U''(ml)=m^2(N-2)\left( N(1+2\beta)-2\right)>0$;\\
	\item $\min_{f>0}\left(\frac{2U}{f^2} \right)=m^2\beta \frac{\beta(1-N)+ (2-N)}{N(1+\beta)-1}<0$.  
\end{itemize}
Hence, it admits $Z_N$-balls according to the criterion (\ref{existence}). 

The shape of potential (\ref{potential1}) has the generic features of Polyakov-loop effective models of SU(N) Yang-Mills theory at finite temperature \cite{Sannino:2005sk}. In our potential, $l$ can be related to the nonzero average norm of the Polyakov-loop at a given temperature $T>T_c$, showing the breaking of $Z_N$-symmetry and deconfinement. In this picture, $f/m$ is the norm of the Polyakov-loop. Note that if $\beta$ goes to a small but negative value, the absolute minimum becomes $f=0$: The case $\beta<0$ would correspond to $T<T_c$. A U(1)-symmetric potential similar to (\ref{potential1}), i.e. $V(\phi^*\phi,(\phi^*\phi)^{N/2})$, has been investigated in \cite{Bazeia:2015gkq}, where analytical formulas for solitons in $(1+1)$-dimensions have been found. The replacement $N\to n+2$ transforms our potential into that of \cite{Bazeia:2015gkq}. 

It is known that the pressure of Yang-Mills matter is given by $p=-\min(V)$, $V$ being the considered effective potential. In our case,
\be
p=-\min(U)=\frac{m^4l^2}{2}\beta\frac{N-2}{N-1}.
\ee
The value of $m^4$ can be matched with the typical pressure scale, that is the Stefan-Boltzmann pressure for a gas of massless gluons. The equality 
\be\label{mdef}
m^4=p_{SB}=\frac{\pi^2(N^2-1)}{45}T^4
\ee
leads to the pressure
\be
p=p_{SB} \frac{N-2}{2(N-1)}l^2\, \beta.
\ee
As shown by lattice simulations, $p<p_{SB}$ for $T\gtrsim T_c$ In the limit $T\to+\infty$, $p\to p_{SB}$ \cite{Borsanyi:2012ve} and $l\to 1$ by definition. It can be concluded that the ``physical" values for $\beta$ are bounded in the range
\be\label{bound}
0<\beta\leq \frac{2(N-1)}{N-2},
\ee
or $0<\beta\leq 4$ at most, the maximal upper bound being reached when $N=3$.

The functions $l(T)$, $\beta(T)$ could be found by fitting the Polyakov loop and pressure computed in lattice QCD, see e.g. \cite{Panero:2009tv} but it is out of the scope of our paper, where we aim at a general discussion of $Z_N$-balls.

\section{\label{sec:num}Numerical results}
We first perform the scalings
\be
f\to m\, l\, f,\qquad r\to \frac{r}{m},
\ee
so that equation (\ref{effpot}) becomes
\be\label{eom:final}
f''+\frac{2}{r}f'-f+(1+\beta)N\, f^{N-1}-\left( N(1+\beta)-1\right)\, f^{2N-3}=0. 
\ee
$\beta$ and $N$ are the remaining parameters. Boundary conditions (\ref{boundary}) are searched for. The three boundary conditions exceed the two conditions associated with the field equation of the second order. The possibility to get non trivial solutions will then consist to fine tune one of the coupling constant (in our case
$\beta$) in function of the control parameter $f_0$. A numerical resolution of the equations (10) can then be performed. We use a collocation method for boundary-value ordinary differential equations, equipped with an adaptive mesh selection procedure \cite{colsys}.

\begin{figure}\label{fig:N3}
	\centering
	\includegraphics[width=0.9\linewidth]{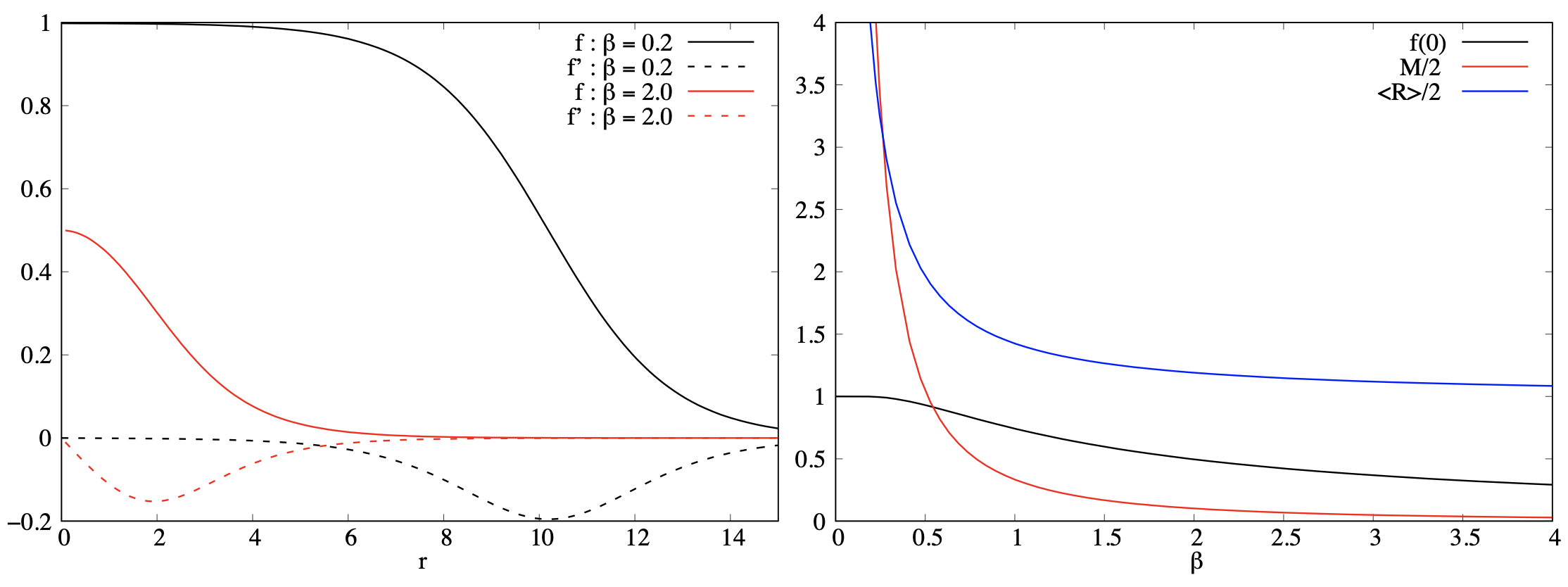}
	\caption{Left panel: $Z_3$-ball profile computed from Eq. (\ref{eom:final}) with $N=3$, $\beta=0.2$ (black lines) or $\beta=2.0$ (red lines). The profile $f$ is displayed (solid lines) as well as its derivative (dashed lines). Right panel: Mass (red line), mean radius (blue line) and value at origin (black line) of the $Z_3$-ball for various values of $\beta$. $f$, $r$ and $M$ are expressed in units of $ml$, $1/m$ and $m$ respectively. }
\end{figure}

\begin{figure}\label{fig:N4}
	\centering
	\includegraphics[width=0.9\linewidth]{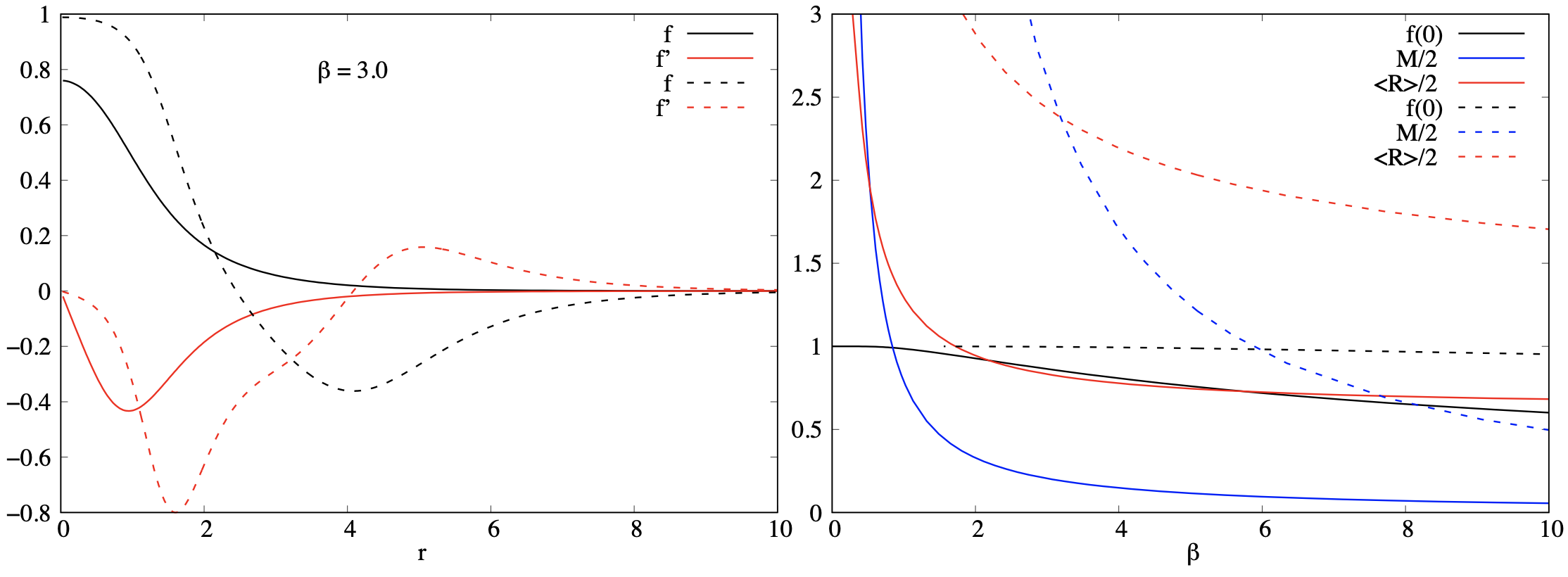}
	\caption{Left panel: $Z_4$-balls profiles computed from Eq. (\ref{eom:final}) with $N=4$ and $\beta=3$. The profile with 0 node and its derivative are displayed (solid lines) as well as the profile with 1 node (dashed lines). Right panel: Mass (blue line), mean radius (red line) and value at origin (black line) of the $Z_4$-ball for various values of $\beta$. $f$, $r$ and $M$ are expressed in units of $ml$, $1/m$ and $m$ respectively. }
\end{figure}

The $Z_3$-ball is displayed in Fig.~2 for various values of $\beta\in[0,4]$ according to the upper bound (\ref{bound}). We notice that $f(0)\to 1$ when $\beta\to 0$: In this case the effective potential is proportional to $f^2(1-f)^2$, namely with two degenerate minima in $f=0$ and $f=1$. The more $\beta$ is small, the more the mean radius is large. At $\beta=0.5$, $M=1.6$ and $\left\langle R\right\rangle=2.7$. The typical energy scale associated with Yang-Mills theory is the deconfinement temperature, $T_c\sim 0.3$ GeV \cite{Lucini:2005vg}. Setting $T=0.3$ GeV and $N=3$ in (\ref{mdef})  leads to $m=345$ MeV and to $M\sim 0.552$ GeV and $\left\langle R\right\rangle\sim 7.83 $ GeV$^{-1}=1.54$ fm. The latter radius is of the same order of magnitude than the profile found in \cite{Gupta:2010pp} with a potential containing the same powers of $\phi$, and the mass is typical of what is observed in  \cite{Brihaye:2020rrp} for temperatures between 1.1 and 1.2 $T_c$. The effective potential reads, for $N=3$ and with the rescaled parameters,
\be \label{potential_N3}
U_{N=3}(f)=\frac{m^4l^2}{2}\left[f^2-2(1+\beta)f^3 +\frac{ 2+3\beta}{2}f^{4}  \right].
\ee
No solution with radial node was found, in agreement with the picture of Fig.~1: The shape of the effective potential when $f<0$ makes impossible for the soliton to reach a zero value at infinity. 

$Z_4$-balls are displayed in Fig.~3 for various values of $\beta$. As for $N=3$, we notice that $f(0)\to 1$ when $\beta\to 0$: In this case the effective potential is proportional to $f^2(1-f^2)^2$, again with two degenerate minima in $f=0$ and $f=1$. Mass and mean radius have similar trends as in the $N=3$ case for the state with zero node. This time, a solution with one radial node has been found. The effective potential reads, for $N=4$ and with the rescaled parameters,
\be \label{potential_N4}
U_{N=4}(f)=\frac{m^4l^2}{2}\left[f^2-2(1+\beta)f^4 +\frac{ 3+4\beta}{3}f^{6}  \right].
\ee
A comparison with Fig.~1 shows that the parity of $U(f)$ indeed allows for radially excited states. We searched for solutions with two radial nodes but did not find any.

\begin{figure}	\label{fig:N}
	\centering
	\includegraphics[width=0.5\linewidth]{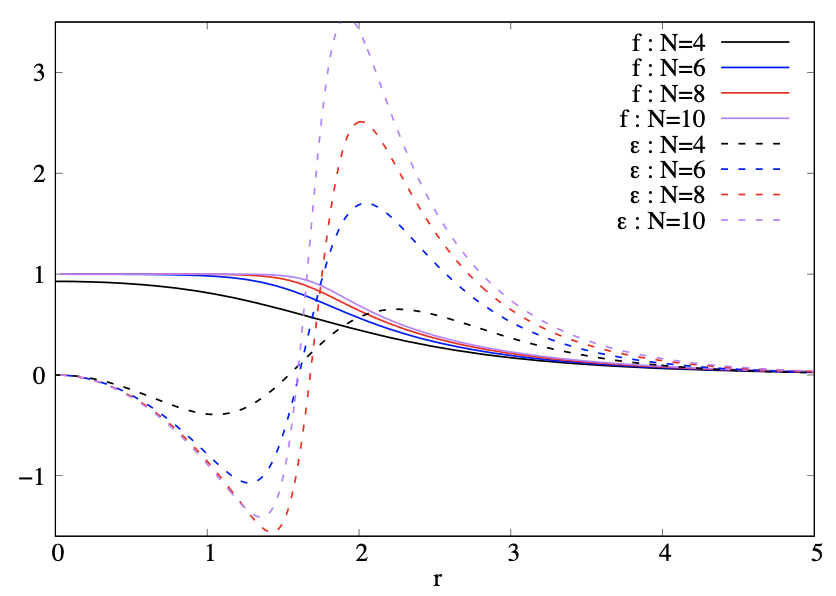}
	\caption{$Z_N$-balls profiles computed from Eq. (\ref{eom:final}) with $\beta=2$ for $N=4$ (black), $N=6$ (blue), $N=8$ (red) and $N=10$ (purple). The profiles with 0 nodes are displayed (solid lines) as well as the local energy density (dashed lines). $f$, $r$ and $\epsilon$ are expressed in units of $ml$, $1/m$ and $m^2$ respectively.}

\end{figure}

The evolution of the zero-node $Z_N$-balls versus $N$ is displayed in Fig.~4. It can be seen that the plateau of the soliton is larger at larger $N$, with an energy density more peaked around its maximal and minimal values. 

\section{\label{sec:solup}Further solutions: Nonconstant phase}
All solutions discussed so far trivialize Eq. (\ref{eom1b}) and the question to deform the $Z_N$ balls
by means of a non-constant phase $\delta({\rm x})$ raises naturally. As a first step
in this direction, it is instructive to study the linearized version of Eq. (\ref{eom1b}) for a radial
phase, i.e. for $\delta(r)$. With the potential (\ref{potential1}) it reduces to ~:

\be
\label{delta_eq}
\delta '' + 2\left(\frac{1}{r} + \frac{f'}{f}\right) \delta' - (1+ \beta) N^2 \delta f^{N-2} = 0 \ \ .
\ee

Using the asymptotic decay (\ref{boundary}) of $f(r)$ in rescaled units where $m=1$, the dominant terms in the asymptotic region reduces to
\be
\label{delta_as}
\delta '' - 2 \delta' = 0 \ \ \ \rightarrow \ \ \ \delta(r\to +\infty) = \delta_0 + \delta_1 e^{2 r} \ \ .
\ee
On the other hand the solutions of Eq. (\ref{delta_eq}) can be expanded around the origin
\begin{eqnarray}
\label{delta_or}
\delta(r) &=& \tilde\delta_0 \left( 1 
+ \frac{N^2 (1+\beta)f_0^{N-2}}{12} r^2 \right. \nonumber\\
&&\left. + \frac{N^2 (1+\beta) f_0^{N-4}\left(N^2 f_0^N (1+\beta) + (6N-20) f_0f''(0)\right)}{480} r^4 \right)\nonumber\\
&&
+ o(r^6),
\end{eqnarray}
where $\tilde\delta_0$ is an arbitrary constant. Note that only the presence of the last term in (\ref{delta_eq}), which in due to the $U(1)$-breaking term, leads  to the  condition (\ref{delta_or}) and  allows for a non-constant and regular $\delta$ at the origin. The U(1) symmetry is recovered for $\beta=-1$ and only allows for a constant $\delta$.

Our numerical integration gave strong evidence that solutions of Eq. (\ref{delta_eq}) extrapolating
between the behaviours (\ref{delta_or}) and (\ref{delta_as}) exist for generic values of $N$ and $\beta$. Solutions of the full system (II.4), if they exist, would be  specific to the $Z_N$ symmetry 
of the potential and  will be addressed in a forthcoming paper.

\section{\label{sec:summary}Concluding comments}

We have shown that spherically symmetric solitons with finite energy may exist in a complex Klein-Gordon model with $Z_N$-symmetric potential. The existence of such $Z_N$-balls in potential $V\left(\phi^*\phi,\phi^N+\phi^{*N}\right)$ is guaranteed provided that $Q$-balls with zero charge exist in the potential $V\left(\phi^*\phi,2(\phi\phi^*)^{N/2}\right)$. As an illustration, we have built solutions in a potential inspired by finite-temperature SU($N$) Yang-Mills theory. We found solitons with zero radial node for all values of $N$, and states with one radial node for even $N$.  In the latter case indeed, the parity of the potential allows for radially excited states.

Our solutions escape Derrick's non existence argument \cite{Derrick:1964ww} because 
our effective potential is non-positive definite for some values of the scalar field. However, the
scaling argument of Derrick leads to an instability. The unstable mode under scaling of the radial variable
can be isolated by perturbating our solution, say $f_0$, according to  $f(r) = f_0(r) + \eta(r)$ and examining
the quadratic term in $\eta$ of the perturbated energy. It turns out that $\eta(r) = f_0'(r)$ constitutes
an unstable mode with eigenvalue $\omega = -2$, independently of $N$ and $\beta$. Keeping in mind the interpretation of $Z_N$-balls as possible ``plasma bubbles" of deconfined Yang-Mills matter, their instability must not be regarded as unphysical since such configurations are expected to decay by nucleation \cite{Biswal:2019xju} and eventually tend to the trivial solution $\phi=0$ while cooling down below $T_c$ \cite{Scavenius:2001pa}.

As an outlook, we mention that more general solutions that those originating in the $Z_N$-ball ansatz we proposed may a priori be found to solve (\ref{eom1})
 Kink solutions have indeed been shown to exist in MSTB model ($Z_2$-symmetry) in $(1+1)$-dimensions since the pioneering works \cite{Montonen:1976yk,SARKER1976255,SUBBASWAMY1981379,Rajaraman:1978kd}. The case $N=3$ has been recently studied \cite{Biswal:2019xju} with the construction of bounce solutions in a $Z_3$-symmetric PNJL model in $(3+1)$-dimensions, i.e. where the Polyakov loop is coupled to a quark matter field. In that spirit, we hope to build solutions of the field equations (\ref{eom1}) with position-dependent phase $\delta$ for arbitrary $N$ in future works. Note that the stability of these solutions is also not guaranteed in more than 1 spatial dimension and will deserve a specific investigation. 

		\begin{acknowledgments}
				We thank Prof. Lukacs Arpad for his comments about the stability of our solutions. 
			\end{acknowledgments}





 %

\end{document}